\documentclass[10pt,A4paper,conference]{IEEEtran}
\usepackage{amsmath}
\usepackage{amssymb}

\begin{document}

\title{A Generalised Hadamard Transform}

\author{\authorblockN{K. J. Horadam}
\authorblockA{Mathematics, SMGS\\
RMIT University\\ Melbourne VIC 3000, Australia \\ Email:
kathy.horadam@ems.rmit.edu.au} }
%

\newtheorem{ex}{Example}[section]
\newtheorem{definition}[ex]{Definition}
\newtheorem{theorem}[ex]{Theorem}
\newtheorem{corollary}[ex]{Corollary}
\newtheorem{lemma}[ex]{Lemma}
\newtheorem{proposition}[ex]{Proposition}

\maketitle

\begin{abstract}
A Generalised Hadamard Transform for multi-phase or multilevel
signals is introduced, which includes the Fourier, Generalised,
Discrete Fourier, Walsh-Hadamard and Reverse Jacket Transforms.
The jacket construction is formalised and shown to admit a tensor
product decomposition. Primary matrices under this decomposition
are identified. New examples of primary jacket matrices of orders
8 and 12 are presented.
\end{abstract}

\section{Introduction}

There are many discrete signal transforms whose transform matrices
have entries on the unit circle. For instance, the Discrete
Fourier Transform (DFT) for signals of length $n$ and the
Walsh-Hadamard Transform (WHT) are both special cases of the
Fourier Transform (FT) found by interpreting the Cooley-Tukey Fast
Fourier Transform in terms of abelian group characters
\cite{MasRoc97}. The family of discrete Generalised Transforms
$\{(GT)_r, 0 \leq r \leq m-1 \}$ for signals of length $2^m$
\cite[10.2]{EllRao82} includes the WHT as case $r = 0$, the
complex BIFORE transform (CBT) as case $r = 1$ and the $2^m$-point
DFT as case $r=m-1$. Both the WHT and DFT are suboptimal discrete
orthogonal transforms, but each has wide application.

These transform matrices all fall into the class of Butson
Hadamard matrices \cite{Butson62}.

For signals where entries outside the complex $n^{th}$ roots of
unity are needed, Lee has introduced multi-phase or multilevel
generalisations of the WHT and of the even-length DFT under the
name {\it Reverse Jacket Transforms} (RJT), see \cite{LeeSRP01}.
These are so-called because the border (``jacket") and the centre
of the unitary matrix representing the transform change
independently under inversion. The former admit a recursive
factorisation into tensor products so represent a fast transform
similar to that of the WHT. (Note that the formula
\cite[Definition 5]{LeeSRP01} is not a reverse jacket transform
generalising both the WHT and even-length DFT, as claimed, because
it is not unitary.)

The class of Butson Hadamard matrices can be generalised to
include all these transform matrices.

\section{A Generalised Hadamard Transform}

In this most general situation, we work in a ring $R$ with unity
1. This includes $\mathbb{R}$, $\mathbb{C}$ and Galois Field
alphabets $GF(p^a)$, though if we need to distinguish signal
values $x$ and $-x$, the ring must have characteristic $\neq 2$.

\begin{definition} \label{def:GBH}
Suppose $R$ is a ring with unity $1$, group of units $R^*$ and
that {\rm char}$\,R$ does not divide $v$. A square matrix $M$ of
order $v \geq 2$, with entries from a subgroup $N \leq R^*$ is a
{\em Generalised Butson Hadamard (GBH)} matrix, if
\[M M^* = M^* M = v I_v,\]
where $M^*$ is the transpose of the matrix of inverse elements of
$M$: $~m_{ij}^* = (m_{ji})^{-1}$. It is denoted {\it GBH}$(N, v)$,
or {\it GBH}$(w, v)$ if $N$ is finite of order $w$.
\end{definition}

A {\it GBH} matrix is always equivalent to a {\it normalised} {\it
GBH} matrix, which has first row and column consisting of all 1s.
By taking the inner product of any non-initial row of a normalised
{\it GBH} matrix $M$ with the all-$1$s first column of $M^*$, we
see that the sum of the entries in any row of $M$, apart from the
first, must equal 0, and similarly for rows of $M^*$ (columns of
the matrix of inverses $M^{(-1)} = [m_{ij}^{-1}]$). The tensor
product of two {\it GBH} matrices over the same group $N$ is a
{\it GBH} matrix over $N$.

\begin{definition}
Let {\bf x} be a signal of length $n$ from $R^*$, where $n \in
R^*$, let $N \leq R^*$ and let $B$ be a {\em GBH}$(N, n)$. A {\em
Generalised Hadamard Transform} (GHT) of ${\bf x}$ is
\begin{equation}
{\bf \hat x} =  B ~{\bf x}
\end{equation} and an {\em Inverse
Generalised Hadamard Transform} (IGHT) of ${\bf \hat x}$ is
\begin{equation}
{\bf x} = n^{-1} B^* ~{\bf \hat x}.
\end{equation}
\end{definition}

The next section describes a construction for {\it GBH} matrices
of even order and additional internal structure, containing the
WHT, even-length DFT and reverse jacket transform matrices.

\section{The jacket matrix construction} \label{ssec:jacket}

Throughout this section, let $G$ be an indexing set of even order
$2n$ (sometimes $G$ is a group such as ${\mathbb Z}_{2n}$ or
${\mathbb Z}_2^a$ but $G$ may be non-abelian or the group
structure may be irrelevant).

\begin{definition} \label{def:jacket}
Let $R$ be a ring with unity $1$. A normalised GBH$(N, 2n)$ matrix
$K$ indexed by $G = \{1, \dots, 2n \}$ with entries from $N \leq
R^*$ is a {\em jacket matrix} if it is of the form
\begin{equation} \label{eq:jm}
K = \left[ \begin{array} {rrcrr} 1 & 1 & \ldots & 1 & 1 \\ 1 & * &
\ldots & * & \pm 1 \\ \vdots & \vdots & \vdots & \vdots & \vdots
\\ 1 & * & \ldots & * & \pm  1 \\ 1 & \pm 1 & \ldots & \pm 1 & \pm
1
\end{array} \right]
\end{equation}
where the central entries $*$ are from $N$. The {\em jacket width}
of $K$ is $m \geq 2$ if $K$ is permutation equivalent to a jacket
matrix $\widetilde{K}$ in which rows $2, \dots, m, 2n-m+1, \dots,
2n$ and columns $2, \dots, m, 2n-m+1, \dots, 2n$ all consist of
$\pm 1$ and $m$ is maximal for this property. Otherwise the jacket
width is $m = 1$.
\end{definition}

All non-initial $\pm 1$ rows and columns of a jacket matrix $K$
necessarily sum to 0 in $R$. If $K$ is a width $m$ jacket matrix
it follows that $K^*$ is itself a width $m$ jacket matrix. If also
$2n \in R^*$, then $K$ has an {\it inverse} $K^{-1} = (2n)^{-1}
K^*$ over $R$. If $(2n)^{\frac{1}{2}} \in R^*$ then $\widehat{K} =
(2n)^{-\frac{1}{2}} K$ is unitary ($\widehat{K} \widehat{K}^* =
I_{2n}$) and we will (by a slight abuse of terminology) also say
$K$ is {\it unitary}.

\begin{ex}
The matrix ${\cal S}_t$ of the WHT of length $2^t$
\begin{equation} \label{eq:WHT}
\begin{array}{cl}
{\cal S}_1 = \left [
\begin{array}{rr}
1 &  1 \\ 1 & -1
\end{array}
\right ] , & {\cal S}_t = \bigotimes^t {\cal S}_1, ~ t \geq 2,
\end{array}
\end{equation}
is an extremal case (jacket width $2^{t-1}$) of such jacket
matrices (``all jacket").
\end{ex}

\begin{ex}
The matrix ${\cal C}_t$ of the CBT of length $2^t$
\begin{equation} \label{eq:CBT}
\begin{array}{l}
{\cal C}_1 = \left [
\begin{array}{rr}
1 &  -i \\ 1 & i
\end{array}
\right ] , \\
{\cal C}_2 = \left[
\begin{array}{rr}
{\cal S}_1 & {\cal S}_1 \\ {\cal C}_1 & -{\cal C}_1
\end{array}
\right] , \\ {\cal C}_m = \left[
\begin{array}{cc}
{\cal C}_{m-1} & {\cal C}_{m-1} \\ {\cal C}_1 \otimes {\cal
S}_{m-2} & -{\cal C}_1 \otimes {\cal S}_{m-2}
\end{array}
\right] ,  m \geq 3.
\end{array}
\end{equation}
is permutation equivalent to a jacket matrix, for $t \geq 2$.
\end{ex}
{\it Proof.} For $t \geq 2$, ${\cal C}_t$ is normalised and the
first two rows of ${\cal C}_t$ consist of $2^{t-1}$ copies of
${\cal S}_1$. By induction the $(2^{t-1}+1)^{st}$ column of ${\cal
C}_t$ is $[ {\bf 1} ~{\bf -1}]^\top$, where ${\bf 1}$ has length
$2^{t-1}$. Rotating the second row to the bottom and the
$(2^{t-1}+1)^{st}$ column to the right of ${\cal C}_t$ produces a
jacket matrix. $~~\square$

\begin{ex}
The matrix ${\cal F}_{2n} = [ ~\omega^{jk} ~]_{0 \leq j, k \leq
2n-1}$, where $\omega = e^{-\pi i/n}$, $n \geq 1$, of the DFT of
length $2n$ is permutation equivalent to a jacket matrix.
\end{ex}
{\it Proof.} \cite[Theorem 1, Definition 1]{LeeSRP01} Represent
the indices in mixed radix notation $j = j_1 n + j_0 = (j_1,
j_0)$, with index set $G = {\mathbb Z}_2 \times {\mathbb Z}_n$.
The permutation $(j_1 , j_0) \mapsto (j_1, ~(1 - j_1)j_0 + (n - 1
- j_0)j_1)$ leaves the first $n$ indices unchanged and reverses
the order of the last $n$ indices. Under this permutation on rows
and columns, ${\cal F}_{2n}$ is equivalent to ${\cal K}_n(\omega)
= $
\begin{equation} \label{eq:jacketDFT}
[ \omega^{\{(1 - j_1)j_0 + (n - 1 - j_0)j_1 + j_1 n \} \{(1 -
k_1)k_0 + (n - 1 - k_0)k_1 + k_1 n \}}],
\end{equation}
the matrix of Lee's complex RJT. $~~\Box$

If $K, K'$ are jacket matrices indexed by $G, G'$ of orders $2n,
2n'$ respectively, with entries from $R^*$, then the tensor
product $K \otimes K'$ is a jacket matrix indexed by $G \times G'$
of order $4nn'$, with entries from $R^*$, since the border
condition is easily seen to be satisfied. In fact, a tensor
product of jacket matrices is a jacket matrix of width $\geq 2$.

\begin{lemma} \label{lem:tensorwidth}
If $K_i$ is a width $m_i$ jacket matrix with entries from $R^*$,
for $i=1,2$, then $K_1 \otimes K_2$ is  a jacket matrix of width
at least $2m_1m_2$.
\end{lemma}
{\it Proof.} Let $K_1$ have order $2n$ and $K_2$ have order $2n'$.
Permute $K_i$ to $\widetilde{K_i}$, so $K_1 \otimes K_2$ is
permutation equivalent to $\widetilde{K_1} \otimes
\widetilde{K_2}$. Let $i \in \{ 1, \dots, m_1, 2n-m_1+1, \dots, 2n
\}$ be an index of an all--($\pm 1$)s row in $\widetilde{K_1}$.
The corresponding $i^{th}$ block row in $\widetilde{K_1} \otimes
\widetilde{K_2}$ consists of $2n$ copies of $\pm 1
\widetilde{K_2}$, so each row indexed $\{2,\dots, m_2, 2n'-m_2+1,
\dots 2n'\}$ of each copy consists of all--($\pm 1$)s,
contributing $2m_2-1$ all--($\pm 1$) rows to the $i^{th}$ block
row of $\widetilde{K_1} \otimes \widetilde{K_2}$. If $i = 1$ the
top row is all 1s and if $i
> 1$ the top row consists of $n$ ${\bf 1}$s and $n$ $-{\bf 1}$s.
Those in the top $m_1$ block rows of $\widetilde{K_1} \otimes
\widetilde{K_2}$ may be permuted to occupy the top $2m_1m_2$ rows
and those in the bottom $m_1$ block rows to the bottom $2m_1m_2$
rows, and similarly for columns. $~~\square$

\section{Tensor decomposition}

If a jacket matrix may be decomposed as a tensor product of two
smaller jacket matrices, the decomposition may be repeated until
no further tensor product decomposition is possible.

\begin{definition}
A jacket matrix of length $2n$ is a {\em primary} jacket matrix
$K_n$ if it is minimal with respect to tensor product, that is,
there are no jacket matrices $K$, $K'$ such that $K_n$ is
permutation equivalent to $K \otimes K'$.
\end{definition}

Examples of primary jacket matrices for $n = 1, \dots, 4$ are
\[
K_1  = {\cal S}_1  ,
\]
\begin{equation} \label{eq:K2}
K_2(r) = \left[ \begin{array} {rrrr} 1 & 1 & 1 & 1 \\ 1 & -r & r &
-1 \\ 1 & r & -r & -1 \\ 1 & -1 & -1 & 1
\end{array} \right] , ~~ r \neq \pm 1 \in R^*
\end{equation}
\begin{equation} \label{eq:K3}
K_3(\alpha) =  \left[ \begin{array} {rrrrrr} 1 & 1 & 1 & 1 & 1 & 1
\\ 1 & \alpha & \alpha^2 & \alpha^5 & \alpha^4  &  -1 \\ 1 &
\alpha^2  & \alpha^4 & \alpha^4 & \alpha^2 & 1  \\ 1 & \alpha^5 &
\alpha^4 & \alpha &  \alpha^2  &  -1 \\ 1 & \alpha^4 & \alpha^2 &
\alpha^2 & \alpha^4 & 1 \\ 1 & -1 & 1 & -1 & 1 & -1
\end{array} \right] ,
\end{equation}
where $ \alpha$ is a primitive $6^{th}$ root of unity in an
integral domain $R$, and
\begin{equation} \label{eq:K4}
K_4(i) = \left[ \begin{array} {rrrrrrrr} 1 & 1 & 1 & 1 & 1 & 1 & 1
& 1 \\ 1 & i & -i & 1 & -1 & i & -i & -1 \\ 1 & -i & -1 & i & i &
-1 & -i & 1 \\ 1 & 1 & i & i & -i & -i & -1 & -1 \\ 1 & -1 & i &
-i & i & -i & 1 & -1 \\ 1 & i & -1 & -i & -i & -1 & i & 1 \\ 1 &
-i & -i & -1 & 1 & i & i & -1  \\ 1 & -1 & 1 & -1 & -1 & 1 & -1 &
1
\end{array} \right] .
\end{equation}

The matrix $K_1$ is the unique $2 \times 2$ jacket matrix, and
$K_2(1) = K_1 \otimes K_1 = {\cal S}_2$, so is not primary. The
matrix $K_2(r)$ for $r \neq \pm 1 \in {\mathbb R}^*$ is a
``centre-weighted Hadamard transform" (CWHT) matrix
\cite{LeeSRP01} and for $r = i \in {\mathbb C}^*$, is ${\cal
K}_2(i)$. The matrix $K_3(\alpha)$ with $ \alpha = e^{i \pi/3}$ is
${\cal K}_3(e^{i \pi/3})$ and with $\alpha$ the fourth power of a
primitive root in $GF(25)$ is an ``extended" complex RJT matrix
\cite[Example 4]{LeeSRP01}. The matrix $K_4(i)$ is described for
the first time here: it is equivalent to the back-circulant matrix
derived from a quadriphase perfect sequence \cite[Example
2]{AraWdL01} of length 8; such sequences are rare.

\begin{corollary} \label{cor:tensorjacket}
A jacket matrix is permutation equivalent to a tensor product of
one or more primary jacket matrices. Conversely, any tensor
product of primary jacket matrices is a jacket matrix.
\end{corollary}

\section{Construction of primary jacket matrices}

By Lemma \ref{lem:tensorwidth} any jacket matrix of width 1 is
primary. The tensor product of two jacket matrices is a jacket
matrix, but in fact, to construct a jacket matrix it is enough
that one factor is a jacket matrix and the other a normalised {\it
GBH} matrix.

\begin{theorem}
Let $B$ be a normalised {\em GBH} matrix and $K$ a jacket matrix,
both with entries in $R^*$. Then $B \otimes K$ is permutation
equivalent to a jacket matrix $(B \otimes K)^\dagger$.
\end{theorem}
{\it Proof.} Let $B$ have order $m$ and $K$ have order $2n$. The
first $2n$ rows of $B \otimes K$ consist of $m$ blocks $K, K,
\ldots, K$, so the $1^{st}$ row is all 1's and the $2n^{th}$ row
is all $\pm 1$s, and similarly for columns. Cyclically permute row
$2n$ to row $2mn$ and row $2n +i$ to row $2n + i-1$, $i = 1,
\ldots , 2n(m - 1)$, and similarly for columns. This shifts row
$2n$ to the bottom of the matrix and column $2n$ to the right of
the matrix, leaving the order of the other rows and columns
otherwise unchanged. This permuted matrix $(B \otimes K)^\dagger$
is of the form (\ref{eq:jm}). $~~\square$

This result explains the generation of some primary jacket
matrices and is fundamental to the construction of Generalised
Hadamard Transforms.

\begin{ex} \label{ex:GBH(3,3)}
Let $\beta \neq 1 \in R^*$ satisfy $\beta^2 + \beta + 1 = 0$, so
$\beta^3 = 1$. Let
\[ B_3 = \left[ \begin{array} {rcc} 1 & 1 & 1
\\ 1 & \beta & \beta^2
\\ 1 & \beta^2 & \beta
\end{array} \right],
\]
so $B_3$ is a normalised GBH$(3, 3)$. Then
\[
(B_3 \otimes K_1)^\dagger =  \left[ \begin{array} {rccccr} 1 & 1 &
1 & 1 & 1 &1 \\  1 &  \beta & \beta & \beta^2 & \beta^2 & 1 \\ 1 &
\beta & -\beta & \beta^2 & -\beta^2 & -1 \\
 1 &  \beta^2 & \beta^2 & \beta & \beta & 1 \\
  1 & \beta^2 & -\beta^2 & \beta & -\beta & -1 \\
  1 & 1 & -1 & 1 & -1  &  -1
\end{array} \right] .
\]
This jacket matrix relates to the DFT matrix of (\ref{eq:K3}) as
follows. A second permutation $(2543)$ cycling central rows and
columns gives the jacket matrix
\[
\left[ \begin{array} {rccccr} 1 & 1 & 1 & 1 & 1 &1
\\  1  &
-\beta & \beta^2 & -\beta^2 & \beta & -1 \\
 1  & \beta^2 & \beta & \beta &  \beta^2 & 1 \\
  1  & -\beta^2 & \beta & -\beta & \beta^2 & -1 \\
1  & \beta & \beta^2 & \beta^2 &  \beta & 1 \\
  1  & -1 & 1 & -1 & 1 &  -1
\end{array} \right] .
\]
When $\alpha$ is a primitive $6^{th}$ root of unity in $R^*$ with
$\alpha^2 = \beta$ and $\alpha^5 = \gamma$, then this matrix
equals ${\cal K}_3(\gamma)$.
\end{ex}

The following matrix is a new $12 \times 12$ primary jacket
matrix, since it has width 1.

\begin{ex}
Let $r \neq \pm 1 \in R^*$.

Let $K_6(\beta, r)$ = $(B_3 \otimes K_2(r))^\dagger$ =

\tiny{
\[
\left[\begin{array} {rrrrrrrrrrrr} 1 & 1 & 1 & 1 & 1 & 1 & 1 & 1 &
1 & 1 & 1 & 1
\\
1 & -r & r & 1 & -r & r & -1 & 1 & -r & r & -1 & -1 \\ 1 & r & -r
& 1 & r & -r & -1 & 1 & r & -r & -1 & -1 \\ 1 & 1 & 1 & \beta &
\beta & \beta & \beta & \beta^2 & \beta^2 & \beta^2 & \beta^2 & 1
\\
1 & -r & r & \beta & -r\beta & r\beta & -\beta & \beta^2 &
-r\beta^2 & r\beta^2 & -\beta^2 & -1 \\ 1 & r & -r & \beta &
r\beta & -r\beta & -\beta & \beta^2 & r\beta^2 & -r\beta^2 &
-\beta^2 & -1 \\ 1 & -1 & -1 & \beta & -\beta & -\beta & \beta &
\beta^2 & -\beta^2 & -\beta^2 & \beta^2 & 1 \\ 1 & 1 & 1 & \beta^2
& \beta^2 & \beta^2 & \beta^2 & \beta & \beta & \beta & \beta & 1
\\
1 & -r & r & \beta^2 & -r\beta^2 & r\beta^2 & -\beta^2 & \beta &
-r\beta & r\beta & -\beta & -1 \\ 1 & r & -r & \beta^2 & r\beta^2
& -r\beta^2 & -\beta^2 & \beta & r\beta & -r\beta & -\beta & -1 \\
1 & -1 & -1 & \beta^2 & -\beta^2 & -\beta^2 & \beta^2 & \beta &
-\beta & -\beta & \beta & 1
\\
 1 & -1 & -1 & 1 & -1 & -1 & 1 & 1 & -1 & -1 & 1 & 1
 \end{array}\right]
\]
}
\end{ex}


Those GHT matrices which are jacket matrices have additional
structure, by virtue of their tensor product decomposition into
primary jacket matrices and their jacket form, which may
particularly suit them to specific applications.

Consider the set of jacket matrices with entries in in an integral
domain $R$
\begin{equation} \label{eq:GJT}
\{K = (\otimes^\ell K_1) \otimes K_2(r)^\epsilon \otimes {\cal
K}_n(\alpha)^\delta ; ~ \ell \geq 0, \epsilon, \delta \in \{0, 1\}
\} ,
\end{equation}
where $r \neq \pm 1 \in R^*$, $\alpha$ is a primitive $2n^{th}$
root of unity and where by $M^0$ we mean the $1 \times 1$ identity
matrix.  When $ \ell \geq 1, \epsilon = 0, \delta = 0$, this is
the WHT. When $ \ell = 0, \epsilon = 0, \delta = 1$ and $R =
{\mathbb C}$ this is equivalent to the $2n$-point DFT. When
$\epsilon = 1, \delta = 0$ and $R = {\mathbb R}$, this is the
CWHT. When $\epsilon = 0, n = 2, \alpha = i \in {\mathbb C}^*,
\delta = 1$, or when $\epsilon = 1, r = i \in {\mathbb C}^*,
\delta = 0$, this is the complex RJT, and when $\epsilon = 0,
\delta = 1$, this is the extended complex RJT.

This uniform classification may make it possible to recognise
common fast algorithms.

Finally, the examples and constructions of primary jacket matrices
above are all instances of a class of matrices called {\it
cocyclic}, investigated by the author and colleagues over the past
decade \cite{Horad00,HorUda00}.

\section{Conclusion}

To summarise: a generalisation of Butson's Hadamard matrices
determines a Generalised Hadamard Transform (GHT). The GHT
includes the Fourier and Generalised Transform families (in
particular the WHT and DFT) and the centre-weighted
Walsh-Hadamard, Complex Reverse Jacket and extended Complex
Reverse Jacket Transforms.

In the jacket case, GHT matrices can be permuted into tensor
products of primary jacket matrices. New primary jacket matrices
may be constructed as tensor products of a Generalised Butson
Hadamard matrix which is not a primary jacket matrix, and a
primary jacket matrix. New examples in orders 8 and 12 have been
given.

Application of the GHT to image processing, error-control coding
and decoding and sequence design are obvious directions for future
research.


\begin{thebibliography}{1}
\bibitem{AraWdL01}
K. T. Arasu and W. de Launey, Two-dimensional perfect quaternary
arrays, {\it IEEE Trans. Inform. Theory} 47(4) (2001) 1482--1493.

\bibitem{Butson62}
A. T. Butson, Generalised Hadamard matrices, {\it Proc. Amer.
Math. Soc.} 13 (1962) 894--898.


\bibitem{EllRao82}
D. F. Elliott and K. R. Rao, Fast Transforms: Algorithms,
Analyses, Applications, Academic Press, New York, 1982.

\bibitem{Horad00}
K. J. Horadam, An introduction to cocyclic generalised Hadamard
matrices, {\it Discrete Appl. Math.} 102 (2000) 115--131.

\bibitem{HorUda00}
K. J. Horadam and P. Udaya, Cocyclic Hadamard codes, {\it IEEE
Trans. Inform. Theory} 46(4) (2000) 1545--1550.



\bibitem{MasRoc97}
D. K. Maslen and D. N. Rockmore, Generalised FFTs - a survey of
some recent results, {\it DIMACS Ser. Discr. Math. Theoret. Comp.
Sci.} 28 (1997) 183--237.

\bibitem{LeeSRP01} Moon Ho Lee, B. Sunder Rajan and
J. Y. Park, A generalized reverse jacket transform, {\it IEEE
Trans. Circuits Syst. II}, 48(7) (2001) 684--690.

\end{thebibliography}
\end{document}